\documentclass[preprint,aps,showkeys,preprintnumbers,amsmath,amssymb,footinbib]{revtex4-2}
\usepackage{mathrsfs}
\usepackage{amssymb}
\usepackage{graphicx}
\usepackage{hyperref}
\usepackage{ulem}
\usepackage{tabularx}
\usepackage{array}
\usepackage{color}

\begin{document}
\title[{Lorentz-covariance of} Position Operator and its Eigenstates for a massive spin $1/2$ field]{{Lorentz-covariance of} Position Operator and its Eigenstates for a massive spin $1/2$ field}

\author{Taeseung Choi}
 \email{tschoi@swu.ac.kr}
\affiliation{Institute of General Education, Seoul Women's University, Seoul 01797, Korea}
\affiliation{School of Computational Sciences, Korea Institute for Advanced Study, Seoul 02455, Korea}
%
%



%
%
%

\begin{abstract}
We present a derivation of a position operator for a massive field with spin $1/2$, expressed in {a representation-independent form} of the Poincar\'e group. Using the recently derived Lorentz-covariant field spin operator, we obtain a corresponding field position operator through the total angular momentum formula. Acting on the Dirac spinor representation, the eigenvalues of the field position operator correspond to the spatial components of the Lorentz-covariant space-time coordinate $4$-vector. We show that the field position operator preserves the particle and the antiparticle character of the states. Thus, the field position operator can serve as a one-particle position operator for both particles and antiparticles, thereby avoiding an unusual fast-oscillating term, known as the Zitterbewegung, associated with the Dirac position operator. We show that the field position operator yields the same velocity as a classical free particle. The eigenstates of the field position operator satisfy the Newton-Wigner locality criteria and transform in a Lorentz-covariant manner. The field position operator becomes particle position and antiparticle position operators when acting on the particle and the antiparticle subspaces, both of which are Hermitian. Additionally, we demonstrate that within the particle subspace of the Dirac spinor space, the field position operator is equivalent to the Newton-Wigner position operator. 
 \end{abstract}



\maketitle

 \section{ Introduction}
Relativistic quantum effects are gaining importance in a variety of disciplines. These include quantum information \cite{TernoRMP}, quantum chaos \cite{Huang2018}, electron beams \cite{OurVortex}, and quantum chemistry \cite{Liu2020}. While non-relativistic quantum mechanics (QM) has offered successful descriptions of many phenomena, it is increasingly apparent that relativistic QM can provide new insights into the quantum physics of the relativistic regime. Quantum field theory reconciles QM with special relativity but faces challenges in accounting for position-dependent quantum effects, such as interference in double-slit experiments and heavy ion collisions \cite{Berdermann}.  

Since the inception of Dirac's work in 1928 \cite{Dirac1928}, numerous efforts have been made to extend classical and non-relativistic quantum mechanical concepts of position into a coherent framework in relativistic quantum mechanics \cite{Pryce48,Moller,NW49, FW50,Fleming64,Choi2015,SilenkoPRA20}. {These approaches have largely focused on Lorentz or Poincar\'e symmetries of the space-time coordinate. Recently, an alternative approach was proposed that departs from traditional Poincar\'e symmetry, instead exploring a different symmetry within the phase space \cite{Kong}. While this new perspective offers interesting insights, our paper concentrates on the conventional approach based on Poincar\'e symmetry.} 

One early candidate for a relativistic position operator was the Dirac (canonical) position operator, which has the same form as the non-relativistic position operator. However, the Dirac position operator fails to satisfy the Heisenberg equation for a free particle \cite{DiracPQM}. In other words, the solution of the Dirac position operator for a free particle exhibits weird high-frequency oscillations, known as the Zitterbewegung, first observed by Schr\"odinger \cite{Schrodinger30}. These oscillations stem from the interference between the positive- and negative-energy solutions of the Dirac equation in the expectation value of the position operator. This suggests that the Dirac position operator is not suitable for a one-particle position operator. 

In the search for a more suitable relativistic position operator, Pryce proposed several definitions of mass center in relativistic mechanics as a candidate for a relativistic position operator, going beyond the concept of a point particle \cite{Pryce48}. Among Pryce's operators, Pryce (e) position operator stands out with commuting components. With a similar approach, M$\o$ller studied a center of mass for a finite closed system and showed that his center of mass has commuting components and does not give rise to Zitterbewegung when applied to the Dirac theory of electrons \cite{Moller}. Interestingly, the position operator of M$\o$ller is essentially equivalent to Pryce (e) position operator. The only difference between the two operators lies in a factor of $2$ in the spin-dependent terms, arising due to the Thomas precession overlooked in M$\o$ller's derivation \cite{oconell}. 

Foldy and Wouthuysen (FW) introduced a novel position operator that manifests as a single-particle coordinate in a specialized representation—known as the FW representation—where the free Dirac Hamiltonian is diagonalized \cite{FW50}. This formulation naturally eliminates Zitterbewegung. However, in the original Dirac representation, the operator turns nonlocal due to the inclusion of momentum-dependent nonlocal terms. FW interpreted this as representing the center of mass for the particle's nonlocal spread within its Compton wavelength, and they thus termed it the 'FW mean position operator.' Importantly, FW themselves recognized that the FW mean position operator is equivalent to Pryce's (e) position operator. {Recently, Silenko et al. \cite{SilenkoPRA20} considered} the FW mean position operator as a natural generalization of the non-relativistic position operator, given its identical form in the FW representation.

Newton and Wigner (NW) introduced a fresh perspective in the quest for a relativistic position operator \cite{NW49}. They identified localized states for elementary systems with non-zero mass and arbitrary spin, which are orthogonal to any states displaced by translation. From these localized states, NW naturally defined what is now known as the NW position operator. Interestingly, they observed that their operator is not only unique but also identical to Pryce's (e) position operator. Given this equivalence among the NW, Pryce (e), and FW position operators, they are collectively referred to as the NW position operator. At this stage, the NW position operator seems to have gained widespread support as a potentially correct relativistic one-particle position operator.  

Nonetheless, the NW position operator raises two key issues. The first is that its localized eigenstate spreads faster than the speed of light, raising concerns about causality breakdown \cite{Fleming65,Hegerfeldt85}. Although this seems problematic, it is mitigated by the fact that the tails of the localized eigenstates outside the light cone are exponentially small and the effects are asymptotically causal \cite{ColemanQFT, Ruijsenaars81}. The second issue involves the covariance of the localized eigenstates under Lorentz transformations. Notably, NW did not explicitly examine this for the spin $1/2$ case in their original work \cite{NW49}. They did, however, acknowledge that for a scalar field, an eigenstate that is localized for one observer does not appear so for another observer in relative motion. This limitation casts doubt on the operator's applicability across different reference frames.

Fleming proposed a manifestly covariant formalism based on the correspondence principle between classical and quantum mechanics, which naturally allows for the construction of space-time $4$-vector operators. In this framework, the previous position operators, such as the NW operator, serve as the spatial components of a covariantly generalized space-time $4$-vector operator under Lorentz transformations \cite{Fleming64}. However, it is important to note that Fleming's formalism diverges from conventional QM, where the $0$th-component of the space-time $4$-vector is a $c$-number used to describe time evolution (with $c = 1$), rather than being an operator.

In this paper, we aim to derive a Lorentz-covariant position operator for the Poincar\'e group for a massive spin $1/2$ field within the conventional formalism of $c$-number time. We anticipate that a position operator corresponding to the Lorentz-covariant field spin operator  \cite{Our21SP} through the total angular momentum of the Poincar\'e group, which we refer to as the field position operator, will meet this criterion. However, constructing a position operator directly from the generators of the Poincar\'e group is not feasible. This differs from the case of the recently derived field spin operator \cite{Our21SP}, whose square is the second Casimir of the Poincar\'e group. 

To define the field position operator, we first compare the total angular momentum of the field position and spin operators with their Dirac counterparts. This process comprises several key steps. First, we separate the orbital angular momentum from the total angular momentum. Next, we impose Fleming's locality condition, requiring that the components of the field position operator commute with each other. Additionally, we impose the condition that the position operator should commute with the spin operator, serving as a rotational generator in an internal space. The full representation space of the Poincar\'e group consists of the direct sum of the particle and the antiparticle subspaces. A suitable position operator for one particle should respect this characteristic. Specifically, its action should maintain the particle and antiparticle character of the state. We will demonstrate that, like the field spin operator \cite{Our21SP}, the field position operator fulfills this requirement.

{

 The two conditions—Fleming's locality and the requirement to commute with the field spin operator—do not uniquely determine the field position operator. To establish its uniqueness, we compare its eigenstates in the (covariant) Dirac spinor representation space to the unique localized states obtained using NW's locality criteria \cite{NW49}. We show that these eigenstates transform covariantly and maintain the characteristics of both the particle and the antiparticle under Lorentz transformations.  Notably, we demonstrate that in the positive-energy (particle) subspace, where the NW position operator is defined, the field position operator becomes equivalent to the NW position operator. Additionally, the field position operator is shown to satisfy the quantum-to-classical correspondence of the velocity operator, thereby avoiding Zitterbewegung.

This paper is organized as follows: In Sec. {\ref{sec:DERV}}, we derive the field position operator in {a representation-independent form} from the field spin operator, using the total angular momentum operator expressed by the Dirac spin and Dirac position operators.
In Sec. {\ref{sec:UNHRCV}}, we investigate the uniqueness of the field position operator by considering the localized eigenstates obtained through the NW locality criteria, and examine the Hermiticity, covariance of the position eigenstates under Lorentz transformations{, as well as the relation to the NW position operator.} In Sec. {\ref{sec:VELNZ}}, we demonstrate the quantum-to-classical correspondence for the velocity, showing that there is no Zitterbewegung. Finally, in Sec. {\ref{sec:CON}}, we conclude our paper.


 \section{ Derivation of the field position operator}
\label{sec:DERV}
In this section, we focus on deriving a representation-independent field position operator based on the generators of the Poincar\'e group. According to a core principle of special relativity, the laws of physics must be invariant under coordinate transformations between inertial frames, {described by the Poincar\'e group} $(\Lambda^\mu_{\phantom{\mu}\nu}, \, {d^\mu})$:
\begin{eqnarray}
{x'}^{\mu}= \Lambda^\mu_{\phantom{\mu}\nu}x^\nu + {d^\mu}, 
\end{eqnarray}
where Einstein summation convention is used for Greek indices $\mu \in \{\,0\,,1\,,2\,,3\,\}$. Latin indices $k \in \{\,1\,,2\,,3\,\}$ will denote spatial components. The metric tensor is $g_{\mu\nu}= \mathrm{diag} (+,-,-,-)$. As massive elementary particles with arbitrary spin are described as irreducible representations of the Poincar\'e group \cite{Wigner39}, it becomes necessary to express a position operator in a form that is independent of any specific representation

While a spin operator has been directly derived from the generators of the Poincar\'e group \cite{Our21SP}, as its square is the group's second Casimir, the same approach does not apply to a position operator. Instead, the total angular momentum operator ${\bf J}$, given by
\begin{eqnarray}
\label{eq:TAM}
{\bf J}=\boldsymbol{ \mathcal{X}}\times {\bf P}+\boldsymbol{ \mathcal{S}},
\end{eqnarray} 
serves as the connecting framework for deriving the field position operator ${\bf X}$ by comparing it with different sets of position and spin operators that share the same ${\bf J}$ and ${\bf P}$. In our notation, spatial vectors are represented in boldface, such as $\boldsymbol{\mathcal{X}}$, ${\bf P}$, $\boldsymbol{\mathcal{S}}$, and ${\bf J}$. Capital letters and {hats} are used to indicate operators, for example, $P^\mu$ and the Dirac position operator $\hat{\bf x}$. As was known \cite{SchwartzQFT,Our14}, the irreducible representation space of the Poincar\'e group consists of both particle and antiparticle subspaces. Therefore, a suitable one-particle position operator should respect the particle and antiparticle characteristics, ensuring that its action within a given subspace does not cause states to extend beyond that subspace.   

As demonstrated in our previous work \cite{Our21SP}, the field spin operator ${\bf S}$ has several desirable properties. These include Lorentz covariance and the preservation of both particle and antiparticle features. In the previous work \cite{Our21SP}, we derived the two spin operator ${\bf S}_{\pm}$ 
\begin{subequations}
\begin{eqnarray}
{\bf S}_+ &=& \frac{1}{m^2} \left( {P}^0 {\bf W} -{\bf P} {W}^0 \right) + i \frac{1}{m^2} {\bf W}\times {\bf P}, \\ 
{\bf S}_- &=& \frac{1}{m^2} \left( {P}^0 {\bf W} -{\bf P} {W}^0 \right) - i \frac{1}{m^2} {\bf W}\times {\bf P},
\end{eqnarray}
\end{subequations}
where $P^\mu P_\mu=m^2$. Here the Pauli-Lubanski operator $W^\mu$ is defined as \cite{Lubanski}: 
\begin{eqnarray}
\label{eq:PL}
 W^\mu = \frac{1}{2} \epsilon^{\,\mu\nu\rho\sigma} J_{\nu\rho} P_\sigma,
\end{eqnarray}
 where $\epsilon^{\,\mu\nu\rho\sigma}$ is the $4$-dimensional Levi-Civita $\epsilon^{\,\mu\nu\rho\sigma}$ with $\epsilon^{1230}=1$. The generator of translation is denoted by $P^\mu$, while $J^{\mu\nu}$ is the generator of the (homogeneous) Lorentz group. 

For completeness, we briefly review results related to spin operators as presented in Ref. \cite{Our21SP}. The two spin operators $S^z_+$ and $S^z_-$ provide two inequivalent representations of the restricted Poincar\'e group, each with the eigenstates $\psi_R(p,\lambda)$ and $\psi_L(p,\lambda)$ that satisfy
\begin{subequations}
\begin{eqnarray}
S_+^z \psi_R(p,\lambda) = \lambda \psi_R(p,\lambda) \\
S_-^z \psi_L(p,\lambda) = \lambda \psi_L(p,\lambda),
\end{eqnarray} 
\end{subequations}
where $\lambda$ is the spin eigenvalue in a representation with spin $s$, and it ranges from $-s$ to $s$. The states $\psi_R(p,\lambda)$ and $\psi_L(p,\lambda)$ are the right-handed and left-handed spinors, respectively. Under space inversion, these states are exchanged. Therefore, the space inversion symmetry of the Poincar\'e group requires the direct sum representations of the right-handed and left-handed representations. The irreducible representation space of the full Poincar\'e group involving space inversion, which we will simply denote as the Poincar\'e group, is thus given by:
\begin{eqnarray}
\psi_{+\epsilon} =\left( \begin{array}{c} \psi_L(p,\lambda) \\ \psi_R(p,\lambda) \end{array} \right) \mbox{ and }  \psi_{-\epsilon} =\left( \begin{array}{c} \psi_L(p,\lambda) \\ -\psi_R(p,\lambda) \end{array} \right).
\end{eqnarray}

Therefore, the field spin ${\bf S}$ is the direct sum of ${\bf S}_+$ and ${\bf S}_-$, given by  
\begin{eqnarray}
\label{eq:FSP}
{\bf S} &=& {\bf S}_- \oplus {\bf S}_+= \frac{1}{m^2} \left( {P}^0 {\bf W} -{\bf P} {W}^0 \right) + i\gamma^5 \frac{1}{m^2} {\bf W}\times {\bf P}.
\end{eqnarray}
For brevity, the direct sum of identity operators $\mathcal{I}\oplus \mathcal{I}$ in the first two terms is omitted. The $\gamma^5$ operator is defined as
\begin{eqnarray}
\gamma^5 =\left( \begin{array}{cc} -\mathcal{I} & 0 \\ 0& \mathcal{I} \end{array}\right),
\end{eqnarray}
where the identity operator $\mathcal{I}$ is the identity matrix of dimension $2s+1$ in a representation with spin $s$. 

Specifically for $s=1/2$, the spin eigenstates $\psi_{\pm \epsilon}$ are
\begin{eqnarray}
\label{eq:DSPAP}
\psi_{\pm\epsilon}(p,\lambda) = \left( \begin{array}{c} \psi_L(p,\lambda) \\ \pm \psi_R(p,\lambda) \end{array}\right) = \left( \begin{array}{c} e^{-\boldsymbol{\sigma}\cdot{\boldsymbol{\xi}}/2} \chi_\lambda \\ \pm e^{\boldsymbol{\sigma}\cdot{\boldsymbol{\xi}}/2} \chi_\lambda \end{array}\right) , 
\end{eqnarray}
where $e^{\mp \boldsymbol{\sigma}\cdot\boldsymbol{\xi}/2}$ are the left-handed and right-handed spinor representations of the standard Lorentz boost $\mathcal{L}_{\bf p}$, and $ \boldsymbol{\sigma}\cdot{\boldsymbol{\xi}}=\sigma^k \xi^k$ with Pauli matrices $\sigma^k$. The standard Lorentz boost is defined such that $p^\mu = {\mathcal{L}_{\bf p}}^\mu_{\phantom{\nu}\nu}k^\nu$, where the rest momentum is specified as $(m,{\bf 0})$. Therefore, $\chi_\lambda$ is the 2-dimensional eigenspinor of the $z$-component of the Pauli spin, i.e., $\sigma^z/2$ at rest frame of the momentum satisfying $\chi^\dagger_\lambda \, \chi_\lambda=1$. Here, the rapidity $ {\boldsymbol{\xi}}$ is given as ${\boldsymbol{\xi}}=\frac{{\bf p}}{\lvert {\bf p} \rvert}\tanh^{-1}\frac{\lvert {\bf p} \rvert}{E_{\bf p}}$, where $E_{\bf p}=\sqrt{{\bf p}^2+m^2}$ and $\lvert {\bf p} \rvert =\sqrt{{\bf p}\cdot {\bf p}}$. Hence, the representation space generated by $\psi_{\pm\epsilon}(p,\lambda)$ is the Dirac spinor representation in the usual Weyl basis expressed as:
\begin{eqnarray}
\label{eq:WEYL}
\gamma^0=\left( \begin{array}{cc} 0 & \mathcal{I} \\ \mathcal{I} & 0 \end{array} \right), ~~\gamma^k=\left( \begin{array}{cc} 0 & \sigma^k \\ -\sigma^k & 0 \end{array} \right)
\end{eqnarray}
with $\gamma^5  =i\gamma^0 \gamma^1\gamma^2\gamma^3$ \cite{SchwartzQFT}. Our previous work has demonstrated that the field spin ${\bf S}$ in Eq. (\ref{eq:FSP}) identifies the unique Lorentz-covariant spin operator for massive elementary fields within the framework of the Poincar\'e group.  

In this paper, our primary focus is on massive spin-$1/2$ particles, encompassing all known fundamental massive fermions. The spin-$1/2$ case can serve as a specific foundation for our analysis. The Dirac spinor representation space is partitioned into particle and antiparticle subspaces, for which $\psi_{\pm \epsilon}(p,\lambda)$ are basis states of each subspace. These spinors are the positive-energy solutions of the particle Hamiltonian $H_P$ and the antiparticle Hamiltonian $H_{AP}$, respectively, given by:
\begin{subequations}
\begin{eqnarray}
\label{eq:HAMP}
H_P &=&\gamma^0 \boldsymbol{\gamma} \cdot{\bf p}+m\gamma^0, \\
\label{eq:HAMAP}
H_{AP}&=&\gamma^0 \boldsymbol{\gamma} \cdot{\bf p}-m\gamma^0.
\end{eqnarray}
\end{subequations}
Further details can be found in Appendix \ref{AppendixA}. We refer to $\psi_{\pm \epsilon}(p,\lambda)$ as the particle and antiparticle spinors, respectively,

The Dirac position operator $\hat{\bf x}=i\partial/\partial {\bf p}$ and the Dirac spin operator $\boldsymbol{\Sigma}/2 = \boldsymbol{\sigma}\oplus \boldsymbol{\sigma}/2$ in the Dirac spinor representation intertwine the particle and antiparticle subspaces. This indicates that they are not appropriate operators for individual particles in relativistic QM. In this section, we will derive the field position operator by comparing two different expressions for the total angular momentum: one defined in terms of the Dirac particle and spin operators, and another defined in terms of the field position and spin operators. In the following section, we will affirm that this field position operator preserves the particle and antiparticle character of the states. This is demonstrated by showing its equivalence to the the position operators defined under the framework of the NW locality, which is already known to maintain these characteristics.   

To do this, the Dirac spin operator $\boldsymbol{\Sigma}/2$ in the Dirac spinor representation can be expressed in a representation-independent form using the generators of the Poincar\'e group as follows \cite{Our21SP}:
\begin{eqnarray}
\label{eq:WSP}
{\bf S}_W = \frac{1}{m} \left( {\bf W} - \frac{{W}^0 {\bf P}}{m+{P^0}}\right).
\end{eqnarray}
This expression was first derived by Bogolubov et al. \cite{Bogolubov}, also referred to as the Wigner spin in recent works \cite{Terno2016,Haen2022}. By comparing the two expressions of the total angular momentum in Eq. (\ref{eq:TAM})-one using the Dirac position and spin operators $\hat{\bf x}$ and ${\bf S}_W$, as defined in Eq. (\ref{eq:WSP}), and the other using the field position and spin operators ${\bf X}$ and ${\bf S}$, as outlined in Eq. (\ref{eq:FSP})-we derive the following expression for the field orbital angular momentum ${\bf X}\times {\bf P}$:
\begin{eqnarray}
\label{eq:AMR}
{\bf X}\times {\bf P} = \hat{\bf x}\times {\bf P}  + \frac{ ({\bf W}\times {\bf P})\times {\bf P}}{m^2  (m+{P^0})} -i\gamma^5 \frac{1}{m^2} {\bf W}\times {\bf P}. 
\end{eqnarray} 
To satisfy this equation, the field position operator ${\bf X}$ takes the form:
\begin{eqnarray}
\label{eq:FPOV1}
{\bf X} = \hat{\bf x} + \frac{ {\bf W}\times {\bf P}}{m^2  (m+{P^0})} -i\gamma^5 \frac{1}{m^2} {\bf W} + \hat{a} {\bf P},
\end{eqnarray}
where the operator $\hat{a}$ could be an arbitrary function of $P^\mu$ and $W^\mu$ to satisfy ${\hat{a}}{\bf P}\times {\bf P}=0$. 

To determine $\hat{a}$, we use the condition $[X^i,\,S^j]=0$, which reflects the fact that a spin operator is a generator that transforms an internal space independent on the space-time. To evaluate $[X^i,\,S^j]=0$, it is convenient to express $W^\mu$ in the Dirac spinor representation with momentum ${\bf p}$ as \cite{Our21SP}:  
\begin{eqnarray}
\label{eq:PLAF}
W^0 = \frac{\boldsymbol{\Sigma}\cdot{\bf p}}{2}, \,\,\, {\bf W}=\frac{m \boldsymbol{\Sigma}}{2}+\frac{{\bf p}(\boldsymbol{\Sigma}\cdot{\bf p})}{2(m+E_{\bf p})}.
\end{eqnarray}
 Using this specific form, the result of this condition enables us to write $\hat{a}$ in terms of generators: 
\begin{eqnarray}
{\hat{a}} =  i\gamma^5 \frac{W^0}{m^2 P^0}+ \hat{b}.
\end{eqnarray}
Here the operator $\hat{b}$ could be an arbitrary function of $P^\mu$, because $[P^\mu, S^j]=0$. 

To arrive at a more refined form of $\hat{b}$, we employ Fleming's locality condition $[{X}^i, {X}^j]=0$ \cite{Fleming64}. Given the canonical commutation relations for Dirac operators, such as $[\hat{x}^i,\, \hat{x}^j]=0$, $[\hat{x}^i,\, P^j]=i\delta_{ij}$, and $[\hat{x}^i, \Sigma^k]=0$, where $\delta_{ij}$ is Kronecker delta,  this condition can be easily calculated. As a result, we find that the freedom of $\hat{b}$ can be reduced to an arbitrary function of ${\bf P}^2={\bf P}\cdot{\bf P}$, denoted as $b({\bf P}^2)$. However, the exact form of $b({\bf P}^2)$ remains to be determined. This additional degree of freedom within the field position operator ${\bf X}$ satisfies the following commutation relationship:
\begin{eqnarray}
[{X}^i + b({\bf P}^2){P}^i ,\,{X}^j + b({\bf P}^2){P}^j]=0.
\end{eqnarray}

Consequently, the field position operator ${\bf X}$3 is expressed as: 
\begin{eqnarray}
\label{eq:FPOR}
{\bf X} = \hat{\bf x} + \frac{ {\bf W}\times {\bf P}}{m^2  (m+{P^0})} -i\gamma^5 \frac{1}{m^2} {\bf W}+i\gamma^5 \frac{W^0{\bf P}}{m^2 P^0}+ b({\bf P}^2){\bf P}
\end{eqnarray}
It is worth noting that the field position operator $X^k$ in Eq. (\ref{eq:FPOR}) complies with the canonical commutation relation with momentum operator, $[X^i, P^j]=i\delta_{ij}$. The significance of the additional degree of freedom in $b({\bf P}^2)$ will be discussed in the following section.

\section{Properties and Equivalence of Field Position Operator}
\label{sec:UNHRCV}

In this section, we aim to provide a comprehensive exploration of the field position operator $X^k$. We begin by investigating the additional degree of freedom represented by $b({\bf P}^2)$ in Eq. (\ref{eq:FPOR}). Following this, we will discuss the Hermiticity and covariance properties of the operator. Subsequently, we will establish its equivalence with the NW position operator in the particle subspace, as defined by \cite{FW50}. Finally, we will explicitly demonstrate the result of the action of the field position operator on a general spinor in the particle subspace. Through this multi-faceted analysis, we aim to offer a comprehensive understanding of the field position operator and its properties.

\subsection{Uniqueness within unitary transformations}\label{sec:Uniqueness}

The issue of investigating the role of the term $b({\bf P}^2)$ in Eq. (\ref{eq:FPOR}) is closely tied to the uniqueness of the field position operator $X^k$. To ensure the uniqueness of the field position operator, it is instructive to compare the field position operator with the unique NW position operator. This NW position operator naturally emerges from localized states that satisfy the NW locality criteria, as outlined in the work by NW \cite{NW49}. 

The NW locality criteria requires that the original state $\psi({ p},\lambda)$ localized at the origin at time $t=x^0$ $(c=1)$ is orthogonal to the translated state $\tilde{\psi}(p,\lambda)$ with a space-like displacement ${\bf d}\neq {\bf 0}$ at the same time. The orthogonality is given by $\langle \psi \vert \tilde{\psi}\rangle =0$ under the following definition of scalar product:
\begin{eqnarray}
\label{eq:LISP}
 \langle \psi \vert {\phi}\rangle =\sum_\lambda \int dp \, \psi^\dagger({ p},\lambda) {\phi}({ p},\lambda),
\end{eqnarray}
where $dp =  \frac{d^3 {\bf p}}{(2\pi)^3} \frac{m}{E_{\bf p}}$ is the Lorentz invariant integral measure on the mass shell and $d^3 {\bf p}=dp^1 dp^2 dp^3$. 

To investigate the NW locality criteria, we consider the purely space-like translation on the Dirac particle and antiparticle spinors:
{\begin{subequations}
\begin{eqnarray}
\label{eq:ESPPO}
\mathcal{T}({\bf d}) \Psi_{+\epsilon}(p,\lambda) &=& e^{i{\bf p}\cdot{\bf d}} \Psi_{+\epsilon}( p,\lambda)  \\
\label{eq:ESAPPO}
\mathcal{T}({\bf d}) \Psi_{-\epsilon}( p,\lambda) &=& e^{-i{\bf p}\cdot{\bf d}} \Psi_{-\epsilon}( p,\lambda), \mbox{ respectively.}
\end{eqnarray}
\end{subequations}
Here, $\mathcal{T}({\bf d})=e^{i {\bf P}\cdot{\bf d}}$ is the translation operator corresponding to a spatial coordinate shift} by ${\bf d}$, i.e., ${\bf x} \rightarrow {\bf x}+{\bf d}$, with $d^{\,0}=0$. The terms $e^{ i{\bf p}\cdot{\bf d}}$ and $e^{-i{\bf p}\cdot{\bf d}}$ are the momentum representations of {$\mathcal{T}({\bf d})$} for a particle and an antiparticle, respectively. The sign reversal of the exponents in the momentum representations is due to the fact that the operator ${\bf P}$ in {$\mathcal{T}({\bf d})$} yields $ \pm{\bf  p}$ when acting on the Dirac plane wave spinors for a particle and an antiparticle, respectively \cite{SchwartzQFT}.

The NW locality criteria are satisfied for both Dirac particle and antiparticle spinors, that is, $\psi({ p},\lambda)=  \psi_{\pm \epsilon}({ p},\lambda)$, with $\tilde{\psi}(p,\lambda) ={\mathcal{T}({\bf d})}\psi_{\pm \epsilon}(p,\lambda) $, due to the equation 
\begin{eqnarray}
\label{eq:NWLC}
\langle \psi \vert \tilde{\psi}  \rangle = \delta({\bf d})=0 \mbox{ for } {\bf d} \neq {\bf 0}.
\end{eqnarray}
This result uses $\psi_{\pm \epsilon}^\dagger({ p},\lambda) \psi_{\pm \epsilon}({ p},\lambda)= E_{\bf p}/m$ from Eq. (\ref{eq:DSPAP}). Here, $\delta({\bf d})$ denotes the Dirac delta function. This fact implies that the Dirac particle and antiparticle spinors $\psi_{\pm \epsilon}(p,\lambda)$ represent localized states at the origin at time $x^0=0$. 

The states $\psi_{\pm \epsilon}(p,\lambda)$, localized at ${\bf x}$ at $x^0=0$, naturally define the position operators $\mathcal{X}^k_P$ and $\mathcal{X}^k_{AP}$ on a hypersurface of simultaneity at $x^0=0$. This is achieved through the following eigenvalue equations:
\begin{subequations}
\begin{eqnarray}
\label{eq:PPOEVQ}
\mathcal{X}^k_P \, e^{-i{\bf p} \cdot {\bf x}} \psi_{+\epsilon}({ p},\lambda) &=& x^k e^{-i{\bf p} \cdot {\bf x}} \psi_{+\epsilon}({ p},\lambda) \mbox{ for a particle} \\
\label{eq:APPOEVQ}
\mathcal{X}^k_{AP} \, e^{i{\bf p} \cdot {\bf x}} \psi_{-\epsilon}({ p},\lambda) &=& x^k e^{i{\bf p} \cdot {\bf x}} \psi_{-\epsilon}({ p},\lambda) \mbox{ for an antiparticle.}
\end{eqnarray}
\end{subequations}
We refer to $\mathcal{X}^k_P$ and $\mathcal{X}^k_{AP}$, as the local particle and antiparticle position operators, respectively. The local particle and antiparticle position operators preserve the particle and antiparticle characteristic, which is guaranteed by their definitions in Eqs. (\ref{eq:PPOEVQ}) and (\ref{eq:APPOEVQ}). In other words, after the action of $\mathcal{X}^k_{P/AP}$, a particle remains a particle, and an antiparticle remains an antiparticle.  

To evaluate the specific forms of the local particle and antiparticle position operators $\mathcal{X}_{P/AP}^k$, we employ a relationship given in Eq. (\ref{eq:DSPAP}) that connects the Dirac spinors $\psi_{\pm \epsilon}(p,\lambda)$ to their rest frame equivalents. This relationship can be reformulated as:  
\begin{eqnarray}
\label{eq:LTPSPAP}
\psi_{\pm \epsilon}({ p}, \lambda) = M(\mathcal{L}_{\bf p}) \psi_{\pm\epsilon}({ k},\lambda),
\end{eqnarray}
where $\psi_{\pm\epsilon}({ k},\lambda)=\chi_\lambda \oplus \chi_\lambda$. In this equation, $M(\mathcal{L}_{\bf p})$ serves as the $4$-spinor representation of the standard Lorentz boost $\mathcal{L}_{\bf p}$. Specifically, $M(\mathcal{L}_{\bf p})$ is the direct sum of the left-handed and right-handed representations, denoted by $e^{\mp \boldsymbol{\sigma}\cdot\boldsymbol{\xi}/2}$. It is noteworthy to mention that $M(\mathcal{L}_{\bf p})$ corresponds to $U(\mathcal{L}_{ p})$ as described in Ref. \cite{Our21SP}. The explicit form $M(\mathcal{L}_{\bf p})$ is given by:
\begin{eqnarray}
\label{eq:SPSLT}
M(\mathcal{L}_{\bf p})= e^{\gamma^5 \boldsymbol{\Sigma}\cdot{\boldsymbol{\xi}}/2} =\frac{E_{\bf p} +m +\gamma^5 \boldsymbol{\Sigma}\cdot{\bf p}}{\sqrt{2m(E_{\bf p}+m)}}.
\end{eqnarray}

It can be deduced that the momentum representations of the local position operators $\mathcal{X}_{P/AP}^k$ at $x^0=0$ are given by:
\begin{subequations}
\begin{eqnarray}
\label{eq:MRPPO}
\mathcal{X}^k_P &=& M(\mathcal{L}_{\bf p}) (i\partial_{p^k}) M^{-1}(\mathcal{L}_{\bf p}), \\
\label{eq:MRAPPO}
\mathcal{X}^k_{AP} &=& M(\mathcal{L}_{\bf p}) (-i\partial_{p^k}) M^{-1}(\mathcal{L}_{\bf p}), \mbox{ respectively},
\end{eqnarray}
\end{subequations}
to satisfy the eigenvalue equations (\ref{eq:PPOEVQ}) and (\ref{eq:APPOEVQ}), using the relations in Eq. (\ref{eq:LTPSPAP}). Note that for both the particle and the antiparticle, two eigenstates exist, corresponding to the spin index $\lambda=\pm1/2$, and both share the same eigenvalue $x^k$.   

The eigenstates of the local position operators $\mathcal{X}^k_P$ and $\mathcal{X}^k_{AP}$ are uniquely determined according to the NW requirements of natural invariance \cite{NW49}. This uniqueness is defined within a unitary transformation applied to the position eigenstates \cite{ColemanQFT}, as the phase factor of the eigenstates is physically irrelevant. Applying a unitary transformation $e^{iB({\bf p}^2)}$ to the position eigenstates $e^{-i{\bf p}\cdot{\bf x}}\psi_{+\epsilon}(p,\lambda)$ for a particle and $e^{i{\bf p}\cdot{\bf x}}\psi_{-\epsilon}(p,\lambda)$ for an antiparticle introduces additional terms to the momentum representations of the local position operators $\mathcal{X}^k_P$ and $\mathcal{X}^k_{AP}$ in Eqs. (\ref{eq:MRPPO}) and (\ref{eq:MRAPPO}) as follows: 
\begin{eqnarray}
\label{eq:UTM}
\mp \frac{\partial}{\partial p^k} B({\bf p}^2) \mbox{ for a particle and an antiparticle, respectively.}
\end{eqnarray}
In this context, $B({\bf p}^2)$ is an arbitrary real function of ${\bf p}^2$.

 To compare the additional term $\mp \partial / \partial_{p^k} B({\bf p}^2)$ in the local position operators to $b({\bf P}^2) P^k$ in the field position operator, we replace the generators $W^\mu$ and $P^\mu$ in the field position operator $X^k$, as given in Eq. (\ref{eq:FPOR}), with their representations in the frame of momentum $p^\mu$. Specifically, we use $W^\mu$ from Eq. (\ref{eq:PLAF}) and $p^\mu=(E_{\bf p}, \,{\bf p})$ for a particle. A sign reversal for an antiparticle arises from the change in the signs of ${P}^\mu$ and ${W}^\mu$ when acting on the antiparticle plane wave states. As a result of this calculation, we find the field position operator $X^k$ on the particle and antiparticle subspaces to be as follows:
\begin{subequations}
\begin{eqnarray}
\label{eq:PPO} \nonumber
{ X}^k_P &=& i\partial_{p^k} +  \frac{(\boldsymbol{\Sigma}\times{\bf p})^k}{2m(m+E_{\bf p})} - i\gamma^5\left( \frac{\Sigma^k}{2m} - \frac{\boldsymbol{\Sigma}\cdot{\bf p} p^k}{2mE_{\bf p}(m+E_{\bf p})} \right) + b({\bf p}^2){\bf p}\\ 
 &{=}& {M(\mathcal{L}_{\bf p}) (i\partial_{p^k}) M^{-1}(\mathcal{L}_{\bf p}) + b({\bf p}^2){\bf p},} \\ 
\label{eq:APPO}
{ X}^k_{AP}&=& - {X}^k_{P}.
\end{eqnarray}
\end{subequations}
The operators {${X}^k_{P/AP}$ will henceforth be referred to as the 'particle and antiparticle position operators' to distinguish them from the local particle and antiparticle position operators $\mathcal{X}^k_{P/AP}$.

The term $b({\bf p}^2)$ in Eqs. (\ref{eq:PPO}) and (\ref{eq:APPO}) must be real to ensure the Hermiticity of the particle and antiparticle position operators. Therefore, the $\pm b({\bf p}^2) p^k$ terms in Eqs. (\ref{eq:PPO}) and (\ref{eq:APPO}) can be eliminated by applying the unitary transformation $e^{i B({\bf p}^2)}$ to the position eigenstates for a particle and an antiparticle, provided they satisfy
\begin{eqnarray}
b({\bf p}^2) p^k =\frac{\partial}{\partial {p^k}}B({\bf p}^2),
\end{eqnarray}
as indicated in Eq. (\ref{eq:UTM}). Thus, we can conclude that the field position operator $X^k$ in a representation-independent form is unique and can be expressed as 
\begin{eqnarray}
\label{eq:FPO}
{\bf X} = \hat{\bf x} + \frac{{\bf W}\times{\bf P}}{{m^2  (m+ {P^0})}} -i\gamma^5 \frac{1}{m^2 } {\bf W} +i\gamma^5 \frac{W^0{\bf P}}{m^2 P^0}, 
\end{eqnarray} 
up to terms that can be removed by unitary transformations of the position eigenstate (i.e., within unitary equivalence). It is noteworthy that this field position operator preserve the particle and antiparticle characteristics. This property is confirmed by the fact that $X^k$ reduces to either the particle position operator $X^k_P$ or the antiparticle position operator $X^k_{AP}$, depending on the space $X^k$ acts, and these are equal to their respective counterparts $\mathcal{X}^k_{P/AP}$.

\subsection{Hermiticity and covariance}

The field position operator ${X}^k$ is Hermitian. Specifically, the unique operator ${X}^k$ given in Eq. (\ref{eq:FPO}) becomes the Hermitian parts of the operators $i\partial_{p^k}$ and $-i\partial_{p^k}$, respectively, after acting on the Dirac particle and antiparticle subspaces. These are formulated as follows: 
\begin{subequations}
\begin{eqnarray}
\label{eq:HPPO}
{X}^k_P &=& \frac{1}{2}\left( i\partial_{p^k} + (i\partial_{p^k})^\dagger \right) = M(\mathcal{L}_{\bf p}) (i\partial_{p^k})M^{-1}(\mathcal{L}_{\bf p}) \\
\label{eq:HAPPO}
{X}^k_{AP} &=& \frac{1}{2}\left( -i\partial_{p^k} + (-i\partial_{p^k})^\dagger \right) = M(\mathcal{L}_{\bf p}) (-i\partial_{p^k})M^{-1}(\mathcal{L}_{\bf p}).
\end{eqnarray} 
\end{subequations}
For position eigenstates $\psi(p,\lambda)=e^{\mp i{\bf p}\cdot {\bf x}}\psi_{\pm \epsilon}(p,\lambda)$, one can directly verify that 
\begin{eqnarray}
\langle \psi \vert X^k \vert \psi \rangle = \langle \psi \vert {X^\dagger}^k \vert \psi \rangle = x^k
\end{eqnarray}
using the scalar product defined in Eq. (\ref{eq:LISP}). It is important to note that the operators $i\partial_{p^k}$ and $-i\partial_{p^k}$ are not Hermitian under this scalar product, which is consistent with the case of scalar particles in NW \cite{NW49}. 

To demonstrate the covariant nature of the eigenstates under Lorentz transformations, we examine the position eigenstates, which are defined on a hypersurface of simultaneity at an arbitrary time slice, $t=x^0$. We obtain the position eigenstates at an arbitrary time $x^0$ from the position eigenstates, $\psi_{\pm \epsilon}(p,\lambda)$, defined at the origin ${\bf x} = {\bf 0}$ and time $x^0 = 0$. This is achieved by the translation operator $\mathcal{T}(x)$ as follows:
\begin{subequations}
\begin{eqnarray}
\label{eq:ESPPOTT}
\mathcal{T}({ x}) \psi_{+\epsilon}({p},\lambda) &=& e^{i{ p}\cdot{x}} \psi_{+\epsilon}(p,\lambda) \mbox{ for particles,}  \\
\label{eq:ESAPPO}
\mathcal{T}({ x}) \psi_{-\epsilon}({ p},\lambda) &=& e^{-i{ p}\cdot{ x}} \psi_{-\epsilon}(p,\lambda) \mbox{ for antiparticles, }
\end{eqnarray}
\end{subequations}
where $p \cdot x= p^\mu x_\mu$. Then, $e^{ \pm i p \cdot x} \psi_{\pm \epsilon}(p,\lambda)$ are the eigenstates of the particle and antiparticle position operators at time ${x}^0$, $X^k_{P/AP}({x}^0)$, respectively, which are defined as
\begin{eqnarray}
\label{eq:POATT}
X^k_{P/AP}({x}^0) = e^{ \pm i p^0 {x}^0 } X^k_{P/AP} e^{ \mp i p^0 {x}^0}.
\end{eqnarray}

Consider a general Lorentz transformation $\Lambda$ that transforms the 4-momentum $p^\mu$ into $q^\mu=\Lambda^\mu_{\phantom{\mu}\nu}p^\nu$ and the 4-position $x^\mu$ into ${x'}^{\mu}= \Lambda^\mu_{\phantom{\mu}\nu}x^\nu$. We can express the transformation of the position eigenstates at time $x^0$ under $\Lambda$ as follows \cite{Our21SP,SchwartzQFT}: 
\begin{subequations}
\begin{eqnarray}
\label{eq:LTPPES}
e^{i{ p}\cdot{ x}} \psi_{+\epsilon}(p,\lambda) \longrightarrow &\phantom{=}& e^{i { p}\cdot \Lambda^{-1}{ x'}} M(\Lambda) \psi_{+\epsilon}(p,\lambda) \\ \nonumber
&=& \sum_{\lambda'} e^{i{ q}\cdot{ x'}}\mathcal{D}_{\lambda'\lambda}(R) \psi_{+\epsilon}({ q},\lambda'), \mbox{ similarly}\\
\label{eq:LTAPPES}
 e^{-i{ p}\cdot{ x}} \psi_{-\epsilon}(p,\lambda) \longrightarrow &\phantom{=}& \sum_{\lambda'} e^{-i{ q}\cdot{ x'}}\mathcal{D}_{\lambda'\lambda}(R) \psi_{-\epsilon}({ q},\lambda'),
\end{eqnarray}
\end{subequations}
where $M(\Lambda)$ is the spinor representation of $\Lambda$. The above transformations can be explained as follows: $M(\Lambda) \psi_{\pm \epsilon}(p,\lambda)$ is equivalent to $M(\Lambda) M(\mathcal{L}_{\bf p}) \psi_{\pm \epsilon}(k,\lambda)$, which further simplifies as follows \cite{Our21SP}:
\begin{eqnarray}
 M(\mathcal{L}_{\Lambda p)}) M^{-1}(\mathcal{L}_{\Lambda p)}) M(\Lambda) M(\mathcal{L}_{\bf p}) \psi_{\pm \epsilon}(k,\lambda)  = M(\mathcal{L}_{\Lambda p)}) \mathcal{D}_{\lambda'\lambda}(R) \psi_{\pm \epsilon}(k,\lambda),
\end{eqnarray}
where $M(\mathcal{L}_{\Lambda p)}) M^{-1}(\mathcal{L}_{\Lambda p)})=I$. Here, $\psi_{\pm \epsilon}(k,\lambda)$ are the spinors in the rest frame of the momentum, as given by Eq. (\ref{eq:LTPSPAP}) and $\mathcal{L}_{\Lambda p}$ is the standard Lorentz boost transforming $k^\mu=(m,{\bf 0})$ into $q^\mu$, because $\Lambda^\mu_{\phantom{\mu}\nu}p^\nu= ({\mathcal{L}_{\Lambda p}})^\mu_{\phantom{\mu}\alpha} R^\alpha_{\phantom{\mu}\nu} k^\nu=({\mathcal{L}_{\Lambda p}})^\mu_{\phantom{\mu}\alpha}k^\alpha$. In these equations, $R=L^{-1}_{\Lambda {\bf p}}\Lambda L_{\bf p}$ represents the Wigner rotation. $\mathcal{D}_{\lambda'\lambda}(R)$ serves as the Dirac spinor representation of the Wigner rotation $R$ in the Weyl basis, which is the 4-dimensional block diagonal matrix. This matrix $\mathcal{D}_{\lambda'\lambda}(R)$ is the direct sum of the standard 2-dimensional rotation matrix of the $SU(2)$ group. Each of these 2-dimensional matrices represents the Wigner rotation $R$ in the particle and antiparticle subspaces, respectively \cite{Our21SP}. It is worth noting that this Wigner rotation of the spinor is responsible for the change of the spin entanglement of two massive Dirac particles under relativistic motion of an observer \cite{Gingrich02}.  

Through Eqs. (\ref{eq:LTPPES}) and (\ref{eq:LTAPPES}), we can see that the Lorentz transformation of the eigenstates of the field position operator in one frame results in the Wigner rotated position eigenstates in the transformed frame, which also satisfy the following Lorentz-covariant eigenvalue equations: 
\begin{subequations}
\begin{eqnarray}
X^k_{P}({x'}^0)  e^{ i{ q}\cdot{ x'}}\mathcal{D}_{\lambda'\lambda}(R) \psi_{+\epsilon}({ q},\lambda') &=& {x'}^k  e^{+ i{ q}\cdot{ x'}}\mathcal{D}_{\lambda'\lambda}(R) \psi_{+\epsilon}({ q},\lambda')\\
X^k_{AP}({x'}^0)  e^{ -i{ q}\cdot{ x'}}\mathcal{D}_{\lambda'\lambda}(R) \psi_{-\epsilon}({ q},\lambda') &=& {x'}^k  e^{- i{ q}\cdot{ x'}}\mathcal{D}_{\lambda'\lambda}(R) \psi_{-\epsilon}({ q},\lambda').
\end{eqnarray} 
\end{subequations}
For simplicity, we have omitted $\sum_{\lambda'}$, as it does not affect the resulting eigenvalue equations. This indicates that the Lorentz-transformed position eigenstates are also eigenstates of the particle and antiparticle position operators $X^k_{P/AP}({x'}^0)$, with a new eigenvalue of ${x'}^k$ at a new time ${x'}^{\,0}$. Consequently, the position eigenvalue and time component, as 4-components of the space-time, undergo a Lorentz transformation according to the Lorentz transformation of the space-time 4-coordinate, while the form of the position eigenstate remains unchanged in the new frame of reference, thereby providing the NW locality criteria, as paralleled in Eq. (\ref{eq:NWLC}). This shows that both the eigenstates and the eigenvalues of the particle and antiparticle position operators—which collectively can be called the field position operator—can be interpreted to undergo Lorentz-covariant transformations. This covariant behavior is desirable for a relativistic position operator. This is because the eigenstates and eigenvalues retain their form in every frame of reference. 

\subsection{Equivalence to the NW position operator}
\label{sec:NWEQ}

Now, we investigate the relationship between the field position operator and the NW position operator for $s=1/2$. The NW position operator was defined on the positive-energy subspace \cite{FW50}, which is equivalent to the particle subspace in the Dirac spinor representation. Therefore, the NW position operator should be compared to the particle position operator $X^k_P$, which serves as the field position operator $X^k$ on the particle subspace. To compare the two position operators, it suffices to consider the operators at time $x^0=0$, as the position operator at time $x^0$ is simply a time translation of the position operator at time $0$ as described in Eq. (\ref{eq:POATT}). 

For this comparison, we rewrite the Dirac particle spinor in Eq. (\ref{eq:LTPSPAP}) by using the unitary operator $U_P({\bf p})$ as follows: 
\begin{eqnarray}
\label{eq:FWTR}
\psi_{+\epsilon}(p,\lambda) = \sqrt{\frac{E_{\bf p}}{m}} U_P({\bf p}) \psi_{+\epsilon}(k,\lambda),
\end{eqnarray}
where $U_P({\bf p})$ is defined by:
\begin{eqnarray}
U_P({\bf p}) = e^{-\gamma^0 \gamma^5 \boldsymbol{\Sigma}\cdot{\boldsymbol{\xi}}/2}.
\end{eqnarray}
One can see the equivalence of $U_P({\bf p})$ to the adjoint of the FW transformation operator in Ref. \cite{FW50}, that is, $U_P({\bf p})=U^\dagger_{FW}({\bf p})$.
During the derivation of the equation (\ref{eq:FWTR}), we utilized:
\begin{eqnarray}
\label{eq:LTFW}
M(\mathcal{L}_{\bf p}) \frac{1+\gamma^0}{2}=\sqrt{\frac{E_{\bf p}}{m}}  U_P({\bf p})  \frac{1+\gamma^0}{2}.
\end{eqnarray}

Consequently, the action of the particle position operator $X^k_P$ in Eq. (\ref{eq:HPPO}) on the eigenstate $e^{- i{\bf p}\cdot {\bf x}}\psi_{+\epsilon}(p,\lambda)$ becomes:
\begin{eqnarray}
X^k_P\, e^{- i{\bf p}\cdot {\bf x}}\psi_{+\epsilon}(p,\lambda) &=& U_P({\bf p}) \left(i \partial_{p^k} -\frac{i}{2}\frac{p^k}{E^2_{\bf p}} \right)U^\dagger_P({\bf p})\, e^{- i{\bf p}\cdot {\bf x}}\psi_{+\epsilon}(p,\lambda) \\ \nonumber
&=&
U^\dagger_{FW}({\bf p}) \left(i \partial_{p^k} -\frac{i}{2}\frac{p^k}{E^2_{\bf p}} \right) U_{FW}({\bf p})\, e^{- i{\bf p}\cdot {\bf x}}\psi_{+\epsilon}(p,\lambda), 
\end{eqnarray}
which corresponds to the FW mean position operator in the Dirac spinor representation in Ref. \cite{FW50}, i.e., the NW position operator for $s=1/2$ \cite{NW49}. The additional term $-{i}{p^k}/(2{E^2_{\bf p}})$ to $i\partial_{p^k}$, which is the same as in the scalar case of NW \cite{NW49}, is due to the extra factor $\sqrt{E_{\bf p}/m}$ in the description by $ U_P({\bf p})$, as described in Eq. (\ref{eq:LTFW}). The additional term $-{i}{p^k}/(2{E^2_{\bf p}})$ is essential for making the position operator Hermitian under the scalar product in Eq. (\ref{eq:LISP}). Therefore, the field position operator $X^k$ is exactly equivalent to the NW position operator when acting on the particle subspace. This exact equivalence is significant because it demonstrates that our derived field position operator in the particle subspace corresponds to the unique localized position operator with commuting components, which is the NW position operator.

Lastly, we close this section by considering the action of the particle position operator $X^k_P$ on a (particle) spinor field $\psi_P({\bf x})$ at time $0$, which is represented as follows:
\begin{eqnarray}
\label{eq:DSF}
\psi_P({\bf x})=  \int dp \sum_\lambda  e^{i{\bf p}\cdot {\bf x}} \psi_{+\epsilon}( p,\lambda) a({ p},\lambda),
\end{eqnarray}
where $a({ p},\lambda)$ is an expansion coefficient satisfying $\int dp \sum_\lambda \lvert a(p,\lambda) \rvert^2 =1$ for a normalized spinor field with $\int d^3{\bf x}\, \psi^\dagger_P({\bf x})\psi_P({\bf x})=1$. Then, we obtain the action of the particle position operator $X^k_P$ on the spinor field $\psi_P({\bf x})$ as:
\begin{eqnarray}
\label{eq:APOSPX}
{X}^k_P \psi_P({\bf x})  = \int dp \sum_\lambda e^{i{\bf p}\cdot {\bf x}} M(\mathcal{L}_{\bf p}) (i\partial_{p^k})M^{-1}(\mathcal{L}_{\bf p})  \psi_{+\epsilon}( p,\lambda).
\end{eqnarray} 

By using integration by parts, we can rewrite Eq. (\ref{eq:APOSPX}) as:
\begin{eqnarray}
\label{eq:SPRPO}
X^k_P \psi_P({\bf x}) &=& x^k \psi_P({\bf x}) \\ \nonumber
&+& \sum_\lambda \int \frac{d^3p}{(2\pi)^3}  e^{i{\bf p}\cdot{\bf x}} \left(-i\partial_{p^k} \frac{m}{E_{\bf p}}\psi_{+\epsilon}(p,\lambda) \right)a(p,\lambda).
\end{eqnarray}
The first term in Eq. (\ref{eq:SPRPO}) is identical to that of the canonical position operator in non-relativistic QM. To estimate the behavior of the second integral, we approximate $\psi_{+\epsilon}(p,\lambda)$ as $\sqrt{E_{\bf p}/m}$ since the norm of $\psi_{+\epsilon}(p,\lambda)$ is $\sqrt{E_{\bf p}/m}$. This approximation essentially yields the same result as in the scalar field case of NW \cite{NW49}, since the position eigenstate for a scalar particle in NW is $\sqrt{E_{\bf p}}$. 

Consequently, the action of the particle position operator on the spinor field can be approximated as:
\begin{eqnarray}
\label{eq:PONFD}
X^k_P \psi_P({\bf x}) = x^k \psi_P({\bf x}) + \frac{1}{8\pi} \int d^3{\bf y} \frac{e^{-m \lvert {\bf x}-{\bf y} \rvert}}{\lvert {\bf x}-{\bf y}\rvert}\frac{\partial \psi_P({\bf y})}{\partial y^k}.
\end{eqnarray}
Here, we use the inverse-Fourier transformation:
\begin{eqnarray}
\frac{m }{E_{\bf p}} \psi_P(p)=\int d^3 {\bf x} \, e^{-i{\bf p}\cdot{\bf x}} \psi_P({\bf x}),
\end{eqnarray}
with the definition of 
\begin{eqnarray}
\psi_P( p)=\sum_\lambda \psi_{+\epsilon}(p,\lambda)a(p,\lambda).
\end{eqnarray}
Note that the magnitude of the second term in $X^k_P \, \psi_P({\bf x}) $ from Eq. (\ref{eq:PONFD}) is constrained to be at most a few multiples of the Compton wavelength, as referenced in \cite{NW49}.

\section{No Zitterbewegung}
\label{sec:VELNZ}


The field position operator $X^k$ preserves the characteristics of both particles and antiparticles. This preservation ensures that there is no conversion between particle and antiparticle states. As Zitterbewegung arises specifically from the interference between particle and antiparticle states, the absence of such conversion effectively rules out the phenomenon of Zitterbewegung. We will confirm the absence of Zitterbewegung through a direct calculation using the velocity operator defined by the field position operator. 

The velocity operator for a particle and an antiparticle is defined as follows in terms of the field position operator $X^k$ \cite{FW50}: 
\begin{eqnarray}
V^k_{P/AP} =\left( \frac{d X^k}{dt} \right)_{P/AP} \equiv -i[X^k, H_{P/AP}],
\end{eqnarray}
where $H_{P/AP}$ refers to the Hamiltonians for a particle and an antiparticle, as given in Eqs. (\ref{eq:HAMP}) and (\ref{eq:HAMAP}), respectively. When acting on the particle and antiparticle subspaces, these velocity operators are represented in momentum space as:
\begin{eqnarray}
V^k_{P/AP} = -i [X^k_{P/AP}, H_{P/AP}]
\end{eqnarray}
for a particle and an antiparticle, respectively. This is because the actions of $H_{P}$ and $H_{AP}$ within the particle and antiparticle subspaces keep the states within their respective subspaces.  

First, we calculate the particle case. A direct calculation with $X^k_{P} $ in Eq. (\ref{eq:HPPO}) yields the following velocity operator for a particle:
\begin{eqnarray}
\label{eq:VEL}
V^k_P = \frac{E_{\bf p}}{m} \gamma^0 \gamma^k +\gamma^k +i \frac{(\boldsymbol{\Sigma}\times {\bf p})^k}{m}- \frac{(\gamma^0 \boldsymbol{\gamma}\cdot {\bf p})p^k}{mE_{\bf p}(m+E_{\bf p})}(E_{\bf p}-m\gamma^0).
\end{eqnarray}
By adding the following terms 
\begin{eqnarray}
\label{eq:ADDT}
\left[ \frac{\gamma^0 \gamma^k}{m} - \frac{(\gamma^0 \boldsymbol{\gamma}\cdot {\bf p})p^k}{mE_{\bf p}(m+E_{\bf p})}\right] (\gamma^0 \boldsymbol{\gamma}\cdot{\bf p}+m \gamma^0 - E_{\bf p}),
\end{eqnarray}
to Eq. (\ref{eq:VEL}), the momentum representation of the velocity operator for a particle simplifies to:
\begin{eqnarray}
\label{eq:VELCLP}
V^k_{\bf p} =\frac{p^k}{E_{\bf p}}.
\end{eqnarray} 
The term $(\gamma^0 \boldsymbol{\gamma}\cdot{\bf p}+\gamma^0 m - E_{\bf p})$ in Eq. (\ref{eq:ADDT}) satisfies the following Dirac particle equation for a particle spinor: 
\begin{eqnarray}
(\gamma^0 \boldsymbol{\gamma}\cdot{\bf p}+\gamma^0 m - E_{\bf p})\psi_{+\epsilon}(p,\lambda)=0.
\end{eqnarray}
Hence, the additional term does not alter the velocity operator for a particle defined in Eq.(\ref{eq:VEL}). This result is as expected, stemming from the equivalence between the particle position operator and the NW position operator in Section \ref{sec:NWEQ}. 

Similarly, by using $X^k_{AP}$ in Eq. (\ref{eq:APPO}), we obtain the momentum representation of the velocity operator for an antiparticle as:
\begin{eqnarray}
\label{eq:VELCLAP}
V^k_{AP}= -\frac{p^k}{E_{\bf p}}.
\end{eqnarray}
According to Eqs. (\ref{eq:VELCLP}) and (\ref{eq:VELCLAP}), the operator forms of the velocities for a particle and an antiparticle can be expressed as:
\begin{eqnarray}
\hat{V}^k= \frac{P^k}{H_{P/AP}}.
\end{eqnarray}
These velocity operators take the same form as the classical velocity for a relativistic particle. Therefore, our findings confirm that there is no Zitterbewegung for a free spinor field, underscoring quantum-to-classical correspondence.

\section{Conclusion}
\label{sec:CON}

In this paper, we derive a position operator for elementary massive fields with spin $1/2$, which we referred to as the field position operator, under the framework of Poincar\'e symmetry. We obtain a representation-independent form of this operator by leveraging the generator form of the field spin operator and a representation-independent total angular momentum. The specific form of the field position operator is determined by imposing the condition that it commutes with the field spin operator and has components that commute. However, there still remains an additional degree of freedom, which allows for an arbitrary function of ${\bf P}^2$. This redundancy is resolved by recognizing that the localized state under the NW criteria in the Dirac spinor representation of the Poincar\'e group is an eigenstate of the field position operator. As a result, the field position operator is shown to be unique up to unitary equivalence and possesses desirable physical properties, including canonical commutation relations, Hermiticity, and covariance under Lorentz transformations. 

Among its properties,  Lorentz covariance stands out as critical for validating the field position operator as a suitable position operator in relativistic QM. As is known, a major reason for questioning the adequacy of the NW position operator as a proper relativistic position operator is the non-covariance of the position eigenstate of the NW position operator in the scalar case, even though the spin $1/2$ case was not explicitly studied by NW. In the case of the field position operator, a position eigenstate and its corresponding eigenvalue with a time component are Lorentz covariant. This implies that the concept of position is preserved for the field position operator when observed by observers in relative motion, akin to the positions of classical relativistic particles. Furthermore, we demonstrate that the field position operator, when projected onto a particle subspace, is equivalent to the NW position operator.

The field position operator preserves the character of both particle and antiparticle states, thereby eliminating interference terms that lead to Zitterbewegung in the context of the Dirac position operator. Through direct calculations, we demonstrate that the velocity operator, defined via the field position operator for a free spinor field, aligns with the classical velocity of a relativistic particle. This result establishes the quantum-to-classical correspondence for the field position operator. Our work suggests a comprehensive and coherent framework for a position operator applicable to massive fields with spin $1/2$, within the realm of Poincar\'e symmetry.


\acknowledgments
 This work was supported by a research grant from Seoul Women's University (2023-0079).

\appendix

\section{Dirac Hamiltonians} \label{AppendixA}

In this appendix, we derive the Dirac Hamiltonians for a particle and an antiparticle by using the parity-space inversion-operation. The parity operation transforms the left-handed spinor $\psi_L(p,\lambda)$ to the right-handed spinor $\psi_R(p,\lambda)$ and vice versa. Hence, the parity operation for the spinors $\psi_{\pm\epsilon}(p,\lambda)$ in Eq. (\ref{eq:DSPAP}) can be represented by $\pm\gamma^0$, as shown in Eq. (\ref{eq:WEYL}). Interestingly, the parity operation is also represented by the square of the inverse standard Lorentz boost, denoted as $M^{-2}(\mathcal{L}_{\bf p})$, which becomes:
\begin{eqnarray}
M^{-2}(\mathcal{L}_{\bf p})= e^{-\gamma^5 \boldsymbol{\sigma}\cdot{\bf f}}=\frac{E_{\bf p}}{m}-\frac{\gamma^5 \boldsymbol{\Sigma}\cdot{\bf p}}{m},
\end{eqnarray}
referring to $M(\mathcal{L}_{\bf p})$ in Eq. (\ref{eq:SPSLT}). 

By equating the two representations of the parity as
\begin{eqnarray}
M^{-2}(\mathcal{L}_{\bf p}) \psi_{\pm\epsilon}(p,\lambda) =\pm \gamma^0 \psi_{\pm\epsilon}(p,\lambda),
\end{eqnarray}
we can derive
\begin{eqnarray}
\pm m \gamma^0 \psi_{\pm\epsilon}(p,\lambda) =( E_{\bf p} -\gamma^5 \boldsymbol{\Sigma}\cdot{\bf p}) \psi_{\pm\epsilon}(p,\lambda).
\end{eqnarray}
These equations yield the Hamiltonians for a particle and an antiparticle, expressed as: 
\begin{eqnarray}
H_{P/AP} \psi_{\pm\epsilon}(p,\lambda) = \gamma^0 \boldsymbol{\gamma}\cdot {\bf p} \pm m \gamma^0
\end{eqnarray}
by using $\gamma^5 \Sigma^k = \gamma^0 \gamma^k$.



\begin{thebibliography}{99}

\bibitem{TernoRMP} Peres, A., Terno, D. R.: {Quantum information and relativity theory}. {Rev. Mod. Phys.} \textbf{76}, 93 (2004) and references therein

\bibitem{Huang2018} Huang, L., Yu, H.-Y., Grebogi, C., Lai, Y.-C.: {Relativistic Quantum Chaos}. {Phys. Rep.} \textbf{753}, 1 (2018) and references therein 


\bibitem{OurVortex} Han, Y. D., Choi, T., Cho, S. Y.: {Singularity of a relativistic vortex beam and proper relativistic observables}. {Sci. Rep.} {\bf 10}, 7417 (2020) and references therein  

\bibitem{Liu2020} Liu, W.: {Essentials of relativistic quantum chemistry}. {J. Chem. Phys.} \textbf{152}, 180902 (2020)

\bibitem{Berdermann} Schweppe, J. {et al.}: {Observation of a Peak Structure in Positron Spectra from U+Cm Collisions}. {Phys. Rev. Lett.} {\bf 51}, 2261 (1983); Cowan, T. {et al.}: {Anomalous Positron Peaks from Supercritical Collision Systems}. {Phys. Rev. Lett.} {\bf 54}, 1761 (1985); Berdermann, E. et al.: GSI Annual Report. GSI 91-1, p161



	\bibitem{Dirac1928} Dirac, P. A. M.: {The quantum theory of the electron}. {Proc. R. Soc. Lond. A} \textbf{117}, 610 (1928)

\bibitem{Pryce48} Pryce, M. H. L.: {The mass-centre in the restricted theory of relativity and its connexion with the quantum theory of elementary particles}. {Proc. R. Soc. Lond. A} \textbf{195}, 62 (1948)


\bibitem{Moller} M$\o$ller, C.: {On the definition of the centre of gravity of an arbitrary closed system in the theory of relativity.} Available online: https://www.stp.dias.ie/Communications/DIAS-STP-Communications-005-Moller.pdf (accessed on 22 January 2020).

 %
\bibitem{NW49} Newton, T. D., Wigner, E. P.: {Localized states for elementary systems}. {Rev. Mod. Phys.} \textbf{21}, 400 (1949)

\bibitem{FW50} Foldy, L. L., Wouthuysen, S. A.: {On the Dirac theory of spin $\mathsf{1/2}$ particles and its non-relativistic limit}. {Phys. Rev.} \textbf{78}, 29 (1950)


%
\bibitem{Fleming64} Fleming, G. N.: {Covariant Position Operators, Spin, and Locality}. {Phys. Rev.} \textbf{137}, B188 (1965)


\bibitem{Choi2015} Choi, T.: {Newton-Wigner position operator and the corresponding spin operator in relativistic quantum mechanics}. {J. Korean Phys. Soc.} \textbf{66}, 877 (2015)


\bibitem{SilenkoPRA20} Zou, L., Zhang, P., Silenko, A. J.: {Position and spin in relativistic quantum mechanics}. {Phys. Rev. A} {\bf 101}, 032117 (2020) and references therein

\bibitem{Kong} Kong, O. C. W., Ting, H. K.:  {$E = mc^2$ versus Symmetry for Lorentz Covariant Physics}, Chinese J. Phys. {\bf 83}, 480 (2023) and references therein


\bibitem{DiracPQM} Dirac, P. A. M.: {The Principles of Quantum Mechanics}. Clarendon, Oxford, (1958)

%
\bibitem{Schrodinger30} Schr\"odinger, E.: {\"Uber die kr\"aftefreie Bewegung in der relativistischen Quantenmechanik}. {Sitz. Preuss. Akad. Wiss. Phys.-Math. Kl.} \textbf{24}, 418-428 (1930)

\bibitem{oconell} O’Connell, R. F.: {Electron interaction with the spin angular momentum of the electromagnetic field}. {J. Phys. A} \textbf{50}, 085306 (2017)


\bibitem{Fleming65} Fleming, G. N.: {Nonlocal Properties of Stable Particles}. {Phys. Rev.} \textbf{139}, B963 (1965)

\bibitem{Hegerfeldt85} Hegerfeldt, G. C.: {Violation of Causality in Relativistic Quantum Theory?}. {Phys. Rev. Lett.} \textbf{54}, 2395 (1985)

%
\bibitem{ColemanQFT} Coleman, S.: {Quantum Field Theory, Lectures of Sidney Coleman}. World Scientific Pub., Hackensack, New Jersey, (2019)

\bibitem{Ruijsenaars81} Ruijsenaars, S. N. M.: {On Newton-Wigner Localization and Superluminal Propagation Speeds}. {Ann. Phys. (NY)} \textbf{137}, 33 (1981)


\bibitem{Our21SP} Choi, T., Han, Y. D.: Lorentz-covariant spin operator for spin $1/2$ massive fields as a physical observable. {J. Korean Phys. Soc.} {\bf 82}, 448–454 (2023)

\bibitem{Wigner39} Wigner, E. P.: {On unitary representations of the inhomogeneous Lorentz group}. {Ann. of Math.} \textbf{40}, 149 (1939); 
Bargmann, V., Wigner, E. P.: {GROUP THEORETICAL DISCUSSION OF RELATIVISTIC WAVE EQUATIONS}. {Proc. Nat. Acad. Sci.} {\bf 34}, 211 (1948)

	
%

%


\bibitem{Our14} Choi, T., Cho, S. Y.: Spin Operators for Massive Particles. arXiv:1410.0468 (2014)

\bibitem{SchwartzQFT} Schwartz, M. D.: {Quantum Field Theory and the Standard Model}. Cambridge University Press, Cambridge, New York, U.S.A. (2013) 


\bibitem{Lubanski} Lubanski, J. K.: {Sur la th\'eorie des particules \'el\'ementaires de spin quelconque}. {Physica (Utrecht)} {\bf 9}, 310 (1942)




\bibitem{Bogolubov} Bogolubov, N. N., Logunov A. A., Todorov, I. T.: {General Principles of Quantum Field Theory}. W. A. Benjamin, Kluwer, Dordrecht, (1990)
	

\bibitem{Terno2016} C\'eleri, L. C., Kiosses, V., Terno, D. K.: {Spin and localization of relativistic fermions and uncertainty relations}. {Phys. Rev. A} {\bf 94}, 062115 (2016)

\bibitem{Haen2022} Lee, H.: {Relativistic massive particle with spin-$1/2$: A vector bundle point of view}. {J. Math. Phys.} \textbf{63}, 012201 (2022)



\bibitem{Gingrich02} Gingrich, R. M., Adami, C.: {Quantum Entanglement of Moving Bodies}. {Phys.Rev.Lett} {\bf 89}, 270402 (2002); Choi, T., Hur, J., KIm, J.: {Relativistic effects on the spin entanglement of two massive Dirac particles}. {Phys. Rev. A} {\bf 84}, 012334 (2012)
	
	
	





\end{thebibliography}
\end{document}